\pdfoutput=1
\documentclass[mnsc]{env/informs3aa} 

\usepackage{natbib}
 \bibpunct[, ]{(}{)}{,}{a}{}{,}%
 \def\bibfont{\small}%
\usepackage[colorlinks=true,bookmarks=false,urlcolor=blue,
citecolor=blue,linkcolor=blue,bookmarksopen=false,draft=false,hypertexnames=false]{hyperref} 
     
\usepackage[dvipsnames]{xcolor}
\usepackage{enumitem}
\usepackage{
amsmath,
amsfonts,
amssymb,
stmaryrd}
\usepackage{mdframed} 
\mdfsetup{backgroundcolor=gray!20}
\usepackage{graphicx}  
\graphicspath{{media/}}
\usepackage{algorithm}
\usepackage[noend]{algpseudocode}
\usepackage[caption=false]{subfig} 
\usepackage[utf8]{inputenc} 
\usepackage[T1]{fontenc}    
\usepackage{url}            
\usepackage{booktabs}       
\usepackage{bbm}
\usepackage{nicefrac}       
\usepackage{multirow, adjustbox}
\usepackage{arydshln}
\usepackage{soul}
\usepackage{bbold} 

\usepackage[skip=0.1em]{caption}
\usepackage{float}
\usepackage{titlesec}
\titlespacing*{\subsection}
{0pt}{0.5em}{0pt}

\usepackage{mathtools}

\usepackage{marvosym}
\usepackage{dsfont}

  {\begin{mdframed}[backgroundcolor=gray!20]}%
  {\end{mdframed}}

\usepackage{tcolorbox} 

\TheoremsNumberedThrough     
\ECRepeatTheorems

\EquationsNumberedThrough    

\MANUSCRIPTNO{MS-0001-1922.65}

 \def\R{\mathbb{R}}

\def\E{\mathbb{E}}

\def\mc{\mathcal}

\def\bF{\boldsymbol}

\def\stemmed{\text{STEMMED}}

\providecommand{\E}{\mathbb{E}}
\providecommand{\R}{\mathbb{R}}

\usepackage{xcolor}

\begin{document}

\pagenumbering{gobble}
{
\SingleSpacedXI


\RUNAUTHOR{Author's names blinded for peer review.}

\RUNTITLE{Improving Opioid Overdose Death Prediction}

\TITLE{Tides Need STEMMED: A Locally Operating Spatio-Temporal Mutually Exciting Point Process with Dynamic Network for Improving Opioid Overdose Death Prediction}

\ARTICLEAUTHORS{%
\AUTHOR{
    Che-Yi Liao\textsuperscript{1}, 
    Zheng Dong\textsuperscript{2}, 
    Gian-Gabriel P. Garcia\textsuperscript{3}, 
    Kamran Paynabar\textsuperscript{1}, 
    Yao Xie\textsuperscript{1}, 
    Mohammad S. Jalali\textsuperscript{4,5}}
    \AFF{
    \textsuperscript{1}H. Milton Stewart School of Industrial and Systems Engineering, Georgia Institute of Technology, Atlanta, Georgia, USA \\ 
    \textsuperscript{2}Amazon.com, Inc, Seattle, WA, USA. \\ 
    \textsuperscript{3}Department of Industrial \& Systems Engineering, University of Washington, Seattle, WA, USA \\ 
    \textsuperscript{4}MGH Institute for Technology Assessment, Harvard Medical School, Boston, MA, USA \\
    \textsuperscript{5}Sloan School of Management, Massachusetts Institute of Technology, Cambridge, MA, USA
    }
}

\ABSTRACT{%
{
\textbf{Problem definition:} Efforts to mitigate the U.S. opioid crisis have been complicated by ever-changing trends in opioid overdose deaths (OODs) across communities and drug types. Public health surveillance efforts are hampered by these challenges, making prediction of local OOD trends and coordination across communities critical. In this research, we design a model-based public health surveillance system capable of leveraging implicit connections between past and future OODs, thereby operating across locales and providing accurate local and global forecasts and unique insights of OOD trends. 
\textbf{Methodology/results:} We develop a Spatio-TEMporal Mutually Exciting point process with Dynamic network (STEMMED) – a point process network wherein each node models a unique community-drug event stream with a dynamic mutually-exciting structure, accounting for influences from other nodes. STEMMED can be decomposed node-by-node, suggesting that it can be tractably parameterized via distributed learning. Leveraging this decomposability, we outline an online cooperative forecasting procedure among local communities and characterize {data-sharing approaches among local entities, including strategies based on drug types, geographical affiliations, and proximity}. We then conduct a numerical study wherein we parameterize STEMMED using individual-level OOD data and city level demographics in Massachusetts. In our offline analysis, we identify a notable cluster formation process of OODs centered around Boston. Additionally, our analysis indicates a growing link between fentanyl and psychostimulants over time. Then, in the online model deployment setting, we find that STEMMED outperforms well-established forecasting models {and that drug-based data sharing policies across cities offer advantages over distance-based county-based and distance-based data sharing policies.} Further, a STEMMED-based OOD surveillance system achieves more than $40\%$ improvement in detection delays relative to evidence-based surveillance systems that are currently hindered by considerable data lags. 
\textbf{Managerial implications:} STEMMED provides accurate forecasts of local OOD trends and highlights complex interactions between OODs across communities and drug types, informing the design and facilitating the timing of impactful policy interventions.
}
}%






}


\maketitle
\newpage
\setcounter{figure}{0}
\setcounter{theorem}{0}
\setcounter{table}{0}
\setcounter{equation}{0}
\setcounter{algorithm}{0}

\pagenumbering{arabic}
\vspace{-1cm}

{

\DoubleSpacedXI 
\section{Introduction}\label{Introduction}

The current opioid overdose crisis has been identified as one of the most severe public health crises in U.S. history \citep{volkow2021changing}. Despite the increasing amount of public and private resources dedicated to preventing opioid overdose deaths (OODs), the situation has remained severe. In 2019, $70\%$ of the $70,000$ drug overdose deaths were associated with opioids, and in 2020, around $65,000$ deaths were identified to be opioid-related, representing $77\%$ of the total drug overdose deaths \citep{mattson2021trends,wonder:2021}. 

Accurate forecasts of overall OOD trends are critical to facilitate effective public health policy planning on both local and national scales \citep{bharat2021big}; however, surveillance efforts have been complicated by the ever-changing drug overdose trends across various geographic regions and population groups \citep{dembek2020opioid, volkow2021changing}. Retrospective analyses have shown that the U.S. has suffered {two or }three waves of the opioid overdose crisis, each having distinctive major causes. The first wave, emerging around 2000, is widely linked to a surge in prescription opioid use and related mortality, though the extent to which this initiated the crisis varies across communities as some local communities had pre-existing heroin epidemics. The second and third waves (which started a decade later) were predominantly driven by illicitly manufactured heroin and fentanyl. More recently, increasing fatalities of psychostimulants-involved OODs have signaled a new wave of the crisis \citep{ciccarone2017fentanyl,ciccarone2021rise, stringfellow2023enumerating, volkow2021changing}. Due to their vastly differing origins, it has been extremely challenging to estimate future OOD progressions, especially with the recent unprecedented presence of fentanyl in the drug supply \citep{stringfellow2023enumerating}. Additionally, detecting new trends from data is challenging because the average lag time (i.e., the time between when the death occurred and when the data are available for analysis) is $4$ to $6$ months. This data recording delay often follows a reporting delay, ranging from a few months to two years across various states \citep{bharat2021big, ahmad2022nchs}. These challenges have stifled policy efforts by the local, state, and federal governments to mitigate the lurking threats of OODs \citep{vadivelu2018opioid}. As such, accurate long-term forecasts of overall OOD trends have become necessary for public health policymaking. 

Moreover, alike other public health surveillance efforts, projecting OOD trends at the local level (e.g., city) could be formidable given the geographically and socioeconomically diverse trends \citep{scholl2019drug, mattson2021trends, liao2025estimating, hsiao2025balancing}. For example, evidence shows that poor and affluent areas are generally concerned with OODs with different drugs \citep{pear2019urban}. These differences are further complicated by racial and socioeconomic disparities in opioid-related morbidity and mortality \citep{lippold2019racial, altekruse2020socioeconomic, liao2022racial,tatar2023social}. Critically, OODs tend to happen in proximity to each other \citep{carter2019spatial,hernandez2020epidemiological}. {This clustering likely arises from a combination of social and structural factors. Social contagion where drug use behaviors and overdose risks spread within networks of individuals who share environments or resources may contribute significantly, as peer influence has been shown to amplify opioid misuse in localized settings \citep{bohnert2011association,carter2019spatial}. Additionally, a shared supply of illicit drugs, particularly those adulterated with potent fentanyl, can heighten overdose risk within a community, given the uneven distribution and unpredictable potency of such substances \citep{Hedegaard2018}. Systemic factors, such as limited access to naloxone or delayed emergency responses in under-resourced areas may further exacerbate these clusters, although their impact varies by region \citep{walley2013opioid}.} The diverse nature of OODs requires customized responses from local and federal health authorities and forecasting approaches that enable cross-community cooperation is essential for addressing the opioid crisis at local and state levels  \citep{jalali2020opioid}. However, how to design such a forecasting system remains an open question. 

To address the need for a systemically designed forecasting model that can be maintained locally, we propose a novel {model-based public health surveillance system centered around a} Spatio-TEMporal Mutually Exciting point process with Dynamic network ($\stemmed$). $\stemmed$ models the dynamic connections between individual OODs via an elegant functional structure that reflects the observations of the space-time clustered OODs. Moreover, $\stemmed$ is coupled with community-level spatial covariates, {e.g., population,} and individual-level socioeconomic information, {e.g., age and sex,} thereby empowering strong short- and long-term forecasting performance. $\stemmed$ can be \textit{localized}, i.e., be decomposed into and operated at various community levels, which enables efficient and effective collective policy design for local public health departments. 

\subsection{Related Works}\label{sec:related_works}
\paragraph{Predictive Models for the Opioid Overdose Crisis.}
{Researchers have modeled OOD progressions using calibrated simulations and statistical learning. \citet{pitt2018modeling} employed a compartmental model to simulate addiction-related OODs, while \citet{homer2021dynamic} incorporated social influence dynamics. More detailed models \citep{Lim2021, stringfellow2022reducing} captured opioid misuse initiation, treatment, and remission but overlooked local heterogeneity, a common gap in opioid modeling \citep{chen2019prevention, linas2021projected}. Statistical approaches have also been explored; \citet{cooper2020modeling} used polynomial regression on time-series data, and \citet{young2018internet, campo2020accurate} leveraged parametric and non-parametric models incorporating web-search data. While promising, these models are sensitive to input features and may struggle to capture OODs' non-stationary space-time dynamics. Additionally, \citet{young2018internet, campo2020accurate} noted privacy concerns and spatial covariate variability, complicating practical implementation.}

\paragraph{Point Process Models with Excitation Structure.} 
One of the important building blocks of $\stemmed$ is the self-exciting point process (SEPPs). These models capture the interactions among a class of event streams and attempt to extract hidden information, termed \textit{triggering effects}, between individual events, which are usually contaminated or lost during data compression or aggregation. Specifically, for event counting processes, these models jointly capture a baseline event rate, independent of the historical events, and total triggering effects, boosted by past events. Due to the rapid development of information technology that allows higher precision and granularity for data collection, point process models with distinctive excitation structures have been applied to various domains, including seismology, criminology, and epidemiology \citep{ogata1988statistical, mohler2011self, chiang2022hawkes}. Importantly, the triggering effects can be modeled on continuous temporal and/or spatio-temporal dimensions with additional features to include more sophisticated effects into the models.  We refer interested readers to \citet{hawkes1971point,hawkes1971spectra,reinhart2018review, hawkes2018hawkes, cheng2025deep} for comprehensive reviews on this topic. 

 For a system of processes, mutually exciting point processes (MEPPs) can be designed to capture relationships among multiple event streams through a network structure, wherein each node corresponds to an event stream and an edge represents the connections between two nodes. Classical MEPPs assume that the edge between a node pair is directed with fixed weights, meaning that the triggering effects for a node incorporate event influences potentially from all nodes in the network.
Researchers have successfully applied classical MEPP to study financial markets, interactions between diverse event streams, such as advertisement clicking and purchases in online shopping investor sentiments, market returns, and invasive species pre- and post-intervention measures \citep{embrechts2011multivariate, xu2014path, yang2018applications, gupta2018discrete}. 

To capture the changing nature of these connections between nodes, dynamic network structures have been investigated using either node-based or edge-based approaches, depending on whether modelers focus on events at nodes or along edges. For node-based methods, \cite{fox2016modeling} studied emails received by each service user, relying solely on emailing frequency between user pairs, while \cite{sanna2024graph} analyzed bike-sharing systems by integrating known graph structures along with event start and end times and nodes. For edge-based approaches, \cite{miscouridou2018modelling} formulated the connections using completely random measures that promote model sparsity. Although it has a relatively small parameter space, this model requires users to fine-tune many hyper-parameters. On the other hand, the study by \cite{sanna2021mutually} focused on modeling events on edges, using an intensity function composed of exponential and Gaussian kernels with node-specific parameters. While this approach avoids the need for fine-tuning, can complicate parameter estimation due to its large parameter space.

Notably, point process models have been applied to OOD-related topics. Without accounting for connections between OOD types, \cite{liu2021point} modeled each OOD stream with emergency calls and OOD-related death reports, whereas \cite{chiang2020sos} treated each node as an OOD category and considered spatio-temporal triggering effects along with static edges between nodes. These studies have established a basis for modeling OODs using triggering effects and highlighted the benefits of integrating heterogeneous data sources and connections across OOD types.

Our approach diverges from existing literature in its modeling objectives and data usage.
$\stemmed$ is equipped with directed edges between nodes to study community-cause-of-death OOD event streams. Unlike the previous works, $\stemmed$ leverages nodal features and individual event features to enhance model precision. Further, $\stemmed$ uses a data-driven parametric form  that combines common drug-use behaviors and discretized spatial effects to replace the incidence matrix usually seen in classical and node-based MEPPs to model the dynamic links between event streams. Importantly, this method allows $\stemmed$ to reduce computational burden of network models and provides in-depth analysis of each nodal process.

\subsection{Contribution}\label{sec:contribution}
 Our contributions are summarized as follows.
\begin{enumerate}
    \item For the applied point process community, we develop a novel data-driven approach for learning point process models with a dynamic network structure, wherein the baseline rate and the individual triggering effects are both modulated by a combination of static and dynamic regressors, and the dynamic connections are highly contextualized.
    $\stemmed$ reduces the parameter space from quadratic to linear by imposing a node-based parameter structure, where estimated parameters depend on individual nodes rather than node-to-node pairs.
    
    \item Within the substance use research community, we are among the first to address complicated global and local system-wide dynamics with one single model, $\stemmed$, and quantify the long-existing phenomenon of space-time clustered OOD events via triggering effects. 
    
    \item For public health practitioners, $\stemmed$ provides an interpretable functional structure that works simultaneously at various community levels, allowing for a timely and accurate estimate and system diagnosis of future OOD trends based only on available data and simple local operations and maintenance.

    {\item Finally, we analyze practical operational strategies when deploying $\stemmed$ as a model-based public health surveillance system, which lays down the foundation for strategic multi-community cooperation.}
\end{enumerate}

The remainder of this paper is structured as follows. In \S \ref{sec: models}, we describe each component of $\stemmed$, including the baseline event rates, the dynamic network connections between unique community-drug event streams, and the individual triggering effect of each OOD event. We also characterize the structure of $\stemmed$, which leads to an efficient node-by-node model-fitting procedure.
{
Next, in \S \ref{sec: deployment}, we outline an online forecasting strategy using $\stemmed$ and discuss key practical concerns and strategies (e.g., data-sharing policies) to facilitate {model operations}.
}
Finally, we present our case study based on illicit OOD forecasting in Massachusetts in \S \ref{sec: case_study} and provide concluding remarks in \S \ref{sec: conclusion}. Additional discussions are relegated to our E-Companion. {\color{red}E-Companion is removed for Arxiv.}
\section{Self/Mutually Exciting Point Process Models and STEMMED} \label{sec: models}
We briefly review preliminary notation for point process models with an excitation structure. We then introduce our Spatio-TEMporal Mutually Exciting point process with Dynamic network (STEMMED) and propose an efficient learning algorithm, which is the key to its localization.

\subsection{Preliminaries on Point Processes}
A point process can be viewed as a set of random vectors whose realizations form a sequence of events: $\bF{x}_1, \bF{x}_2, ...$ where each event $\bF{x}=(t_{\bF{x}}, \bF{m}_{\bF{x}})$ is described by an event time $t_{\bF{x}}\in\R^+$ and mark $\bF{m}_{\bF{x}} \in \mc{M}$. The set $\mc{M}$ represents a \textit{marked space} that contains all the information of an event other than time. In general, one can take $\mc{M}\subset \R^p$ to represent the numerical encoding of additional event information, such as location, drug type, and decedent characteristics. For notational convenience, we define the set $\mc{X}:=\R^+ \times \mc{M}$.

 For any time $t >0$, we define the history of events $\mathcal{D}^{(t)}:= \{\bF{x} \in \mc{X}: t_{\bF{x}} < t\}$ as the set of all information about the process \textit{right before} $t$.  Then, we let the counting measure defined on $\mc{X}$ be ${N}_\mc{S} = |\mc{D}^{(T)} \cap \mc{S}|$, for any time horizon $T>0$ and any measurable set $\mc{S}\subset\mc{X}$. For a function $f:\mc{X} \to \R$, the integral with respect to the counting measure $N_\mc{S}$ can be defined as:  
    $\int_{\mc{S}} f(\bF{x}) d{N_{\bF{x}}}  
        = \sum_{\bF{x}\in N_\mc{S}} f (\bF{x})$.
Importantly, a point process can be characterized by its \textit{conditional intensity function} (CIF) under history $\mc{D}^{(t_{\bF{x}})}$, defined as: $\lambda(\bF{x}\mid\mc{D}^{(t_{\bF{x}})})= \E(d{N_{\bF{x}}} \mid \mc{D}^{(t_{\bF{x}})})/d\bF{x}$. 

In simple words, the CIF represents the \textit{conditional probability} of observing a particular realization at $\bF{x}\in\mc{X}$, given all the observed points up to time $t_{\bF{x}}$. It is common to express $\lambda(\bF{x}\mid\mc{D}^{(t_{\bF{x}})})$ as $\lambda(\bF{x})$ when there is no confusion with the history on which the point process depends.

\subsection{System of Point Processes with Excitation Structure}\label{sec:pp_overview}
The mutually-exciting point process (MEPP), introduced by \citet{hawkes1971spectra}, is a special type of point process that is especially relevant for modeling systems of processes wherein events from one node (a member process) in a network can influence future events in other nodes within the network. Specifically, the CIF of a node $u$ in MEPP is controlled by two components: the baseline
event rate, which is independent of past events, and a total triggering effect, which is collected from the positive influence of every past event from any node in the network, including the node $u$ itself. Suppose that an MEPP system has $M$ nodes, which form a fully connected and directed network. Then, the node $u$ in the MEPP has the following CIF: \begin{equation} \label{eqn: mepp}
    \lambda_u(t) = \mu_u + \sum_{v\in [M]} A_{u}^{v}\sum_{{\bF{x}\in\mc{D}^{(t)}_v}} {\kappa_{u}^v}(t- t_{\bF{x}}),
\end{equation}
where $\mu_u \geq 0$ is the constant background event rate for the node $u$, $A_{u}^{v}\geq 0$ is a constant that controls the magnitude of the triggering effects from node $v$ to node $u$, $\mc{D}^{(t)}_{v}$ is the history of events up to time $t$ at node $v$, and ${\kappa_{u}^v}\geq 0$ is a kernel that specifies the influence of a past event from node $v$ to node $u$. Usually, ${\kappa_{u}^v}(\cdot)$ is assumed to {depend solely on the \textit{time difference} between current time and an event time, rather than the absolute event times}, meaning that
${\kappa_{u}^v}(t-t_{\boldsymbol{x}_1}) = {\kappa_{u}^v}( t - t_{\boldsymbol{x}_2})$ if $t_{\boldsymbol{x}_1} = t_{\boldsymbol{x}_2}$ for some events $\boldsymbol{x}_{1}$ and $\boldsymbol{x}_{2}$.
When the network consists of only a single node (M = 1), the MEPP reduces to a self-exciting point process (SEPP), where events only trigger future events within the same stream.

\subsection{Spatio-Temporal Mutually Exciting Point Process with Dynamic Network} \label{sec: stemmed}
We now detail the formulation for STEMMED. Let $I$ and $S$ be the index sets for local communities and underlying causes of death (i.e., by specific drug types and classes) for OODs, respectively. Then, similar to MEPP, we define STEMMED by a set of nodes representing all community-drug OOD pairs $I\times S$, and arcs, representing the interactions between the nodes. To model the dynamic network arcs and the personalized (i.e., decedent-specific) influence, we extend the basic MEPP form \eqref{eqn: mepp} and formulate each event stream at node $\bF{u} \in I \times S$ as:\begin{equation} 
    \lambda_{\bF{u}}(t) 
    = \mu_{\bF{u}}(t) 
      + \sum_{ \bF{v} \in I\times S} A_{\bF{u}}^{\bF{v}}(t) \sum_{\bF{x}\in\mc{D}_{\bF{v}}^{(t)}}  \kappa_{\bF{u},\bF{x}}(t- t_{\bF{x}})\label{eq: local_stemmed_final}, 
\end{equation}
where $\mu_{\bF{u}}(t) \geq 0$ is the background event rate at time $t$, $A_{\bF{u}}^{\bF{v}}(t) \geq 0$ dynamically controls the magnitude of influences from events occurred at node $\bF{v}$ to node $\bF{u}$, $\kappa_{\bF{u},\bF{x}}(t - t_{\bF{x}} )\geq0$ quantifies the personalized triggering effect from the past event $\bF{x}$ to node $\bF{u}$ at time $t$.  
{
 In the following, we detail \stemmed's model formulation, including the Dynamic Background Event Rate $\left(\mu_{\boldsymbol{u}}(t)\right)$, the Personalized Triggering Effects $\left(\kappa_{\boldsymbol{u}, \boldsymbol{x}}(t-t_{\boldsymbol{x}})\right)$, and the Dynamic Connection $\left( A_{\boldsymbol{u}}^{\boldsymbol{v}}(t) \right)$. We present the complete model in \eqref{eqn:complete_model}} and refer readers to \S \ref{sec:kernel_design_app} in our E-companion regarding further discussions on our model design.

\paragraph{Dynamic Background Event Rate.}
 We consider a practical setting where \textit{nodal-information}, i.e., spatial covariates determining the baseline event rate, are updated in discrete times and possibly asynchronously. For example, one can consider the US Census reports and information on state-wide programs for opioid use disorder treatment. While the former could be released annually, the latter may be published irregularly. Now, assuming $p$-dimensional nodal-information, we model the baseline event rate $\mu(t)$ in a log-linear form, as is often used for point-process-type public health surveillance models \citep{meyer2012space,chiang2022hawkes}: \begin{equation} \label{eq: mu}
    \mu_{\bF{u}} (t) := \gamma_{\bF{u}}\exp\left(
        \bF{\beta_{\bF{u}}}^\top \bF{y}_{\bF{u}, t} 
    \right),
\end{equation}
where $\gamma_{\bF{u}}\in\mathbb{R}^+$ controls the magnitude and $\boldsymbol{\beta}_{\bF{u}} \in \mathbb{R}^p$ is the regression parameter corresponding to the last update of nodal information $\boldsymbol{y}_{\bF{u}, t}$, e.g., the census reports and the treatment program information, \textit{prior} to time $t$.
\paragraph{Personalized Triggering Effects.}
To personalize the effect of each OOD incidence, we consider a general case wherein individual features and event times are both available. We define the personalized temporal kernel $\kappa_{\bF{u},\bF{x}}$ via a product of a \textit{temporal effect} $\kappa_{\bF{u}}$ and an \textit{individual effect magnitude controller} $\eta_{\bF{u},\bF{x}}$: \begin{equation}
\kappa_{\boldsymbol{u},\bF{x}}(t-t_{\bF{x}}) = \eta_{\bF{u},\bF{x}}\kappa_{\bF{u}}(t-t_{\bF{x}}),
\label{eq:stemmed_kappa}
\end{equation}
where the pure temporal effect $\kappa_{\bF{u}}(t-t_{\bF{x}}) = \exp(-\delta_{\bF{u},k}(t-t_{\bF{x}}))>0$ is the exponential kernel adopted from standard SEPPs and MEPPs and is supported by our empirical results.
We then define the {individual effect magnitude controller} $\eta_{\bF{u},\bF{x}}:\mathbb{R}^q\to\mathbb{R}^+$ with a log-linear form, common for personalized effects \citep{meyer2012space, chiang2022hawkes}: \begin{equation} \label{eq: eta}
    \eta_{\bF{u},\bF{x}} = 
        \exp\left(\bF{\omega}^{\top}_{\bF{u}} \bF{m}_{\bF{x}} \right),
\end{equation}
where $\bF{m}_{\bF{x}}$ denotes the $q-$dimensional individual feature vector of $\bF{x}$ in addition to event time, e.g., socioeconomic status and personal medical history, and $\bF{\omega}_{\bF{u}}\in\R^q$ are the coefficients associated with these features.
\paragraph{Dynamic Connections between Event Streams.} To capture the dynamic weights on the network structure in STEMMED, we model $A^{\bF{v}}_{\bF{u}}(t)$ using \textit{discretized spatial effects} and a \textit{social connectivity}.
{

Specifically, we first let nodes $\bF{u}$ and $\bF{v}$ represent community-cause-of-death pairs $(i,s)$ and $(j, s')$, respectively. Then, we define $d_{\bF{u}}^{\bF{v}} \geq 0$ as the physical distance between $\bF{u}$ and $\bF{v}$, i.e., the distance, e.g., miles, between the community centers of $i$ and $j$. Since the spatial coverage of the community set $I$ can be prohibitively huge, we standardize this physical distance for each node $\bF{u}$. Furthermore, we define $\theta_{\bF{u}, t}^{\bF{v}}\in [0,1]$ as the social connectivity between $\bF{u}$ and $\bF{v}$. Intuitively, $\theta_{\bF{u}, t}^{\bF{v}}$ represents the shared drug use behavior between ${\bF{u}}$ and ${\bF{v}}$ at some time $t$, which can be estimated as the percentage of OOD events at community $i$ and $j$ that involved both drugs $s$ and $s'$, prior to time $t$.
} Then, we model the dynamic connections between the nodes in STEMMED as: \begin{equation}\label{eq: R}
    A^{\bF{v}}_{\bF{u}}(t) := \alpha_{\bF{u}} g^{\bF{v}}_{\bF{u}}(t),
    \;\;\text{where}\;\;
    \alpha_{\bF{u}}\geq0,\, g^{\bF{v}}_{\bF{u}} (t) >0,
\end{equation}
with $g^{\bF{v}}_{\bF{u}} (t)$ being a modified spatial kernel capturing the connectivity between $\bF{u}$ and $\bF{v}$ defined as
\begin{equation}\label{eq: g}
    g^{\bF{v}}_{\bF{u}}(t) = \exp{\left(-\delta_{\bF{u}, d}d_{\bF{u}}^{\bF{v}} + \delta_{\bF{u},s} \theta_{\bF{u},t}^{\bF{v}} \right),} 
    \;\;\text{where}\;\;
     \delta_{\bF{u},d},\,\delta_{\bF{u},s}\geq 0.
\end{equation}
Intuitively, each stream $\lambda_{\bF{u}}(t)$ consists of two major components: (1) a time-varying background event rate that is unaffected by the triggering effects, and (2) a total triggering effect, which is a sum of the effects from \textit{every} past event in the system. Notably, unlike MEPP, STEMMED captures the subtle differences in the triggering effects of the events via the personalized temporal kernel $\kappa_{\bF{u},\bF{x}}$. These personalized triggering effects are allowed to be further amplified or attenuated over time as captured by the dynamic connection $A_{\bF{u}}^{\bF{v}}(t)$ and has the added benefit of reducing the number of parameters from quadratic (in a MEPP formulation) to linear (in \stemmed) with respect to the size of the network $|I\times S|$.
{
Notably, our design of the triggering effects can be regarded as a special case of the spatio-temporal kernels in the point process literature \citep{reinhart2018review}. Specifically, our kernel incorporates temporal, spatial, and personalized effects of an event $\boldsymbol{x}$ onto a node $\bF{u}$. Because one of our important aims in this study is to infer actionable policy making at various geographic scales, i.e., state-level and city-level, we model the spatial component as a discretized distance between local communities, which allows us to split the triggering effects into a personalized component and network component. 
Additionally, we remark that STEMMED, similar to other point process models designed for various applications, e.g., \cite{ogata1981lewis, gupta2018discrete,sanna2021mutually,dong2023non,dong2024spatio}, can be regarded as a {\em mechanism-inspired statistical model}. These point processes do not model the exact { causal} mechanism between events but attempt to capture the triggering effects from past to future events from a statistical viewpoint. For STEMMED, we assume that historical OOD incidences may increase the occurrence rate of OODs at other nodes at various magnitudes reflected by the learned kernel. 
}

\subsection{Parameter Estimation}\label{sec: parameter_estimation}
{

Combining~\eqref{eq: local_stemmed_final} -- \eqref{eq: g}, the complete model formulation of $\stemmed$ for node $\bF{u}$ is
\begin{equation}
    \begin{aligned}        
\lambda_{\bF{u}}(t) = \gamma_{\bF{u}}\exp\left(\bF{\beta}_{\bF{u}}^\top \bF{y}_{\bF{u},t}\right)
+ \sum_{\bF{v}{\in I\times S}}\alpha_{\bF{u}}\exp\left(
-\delta_{\bF{u},d} d_{\bF{u}}^{\bF{v}} + \delta_{\bF{u},s}\theta_{\bF{u},t}^{\bF{v}}
\right) \sum_{\bF{x} \in \mathcal{D}_{\bF{v}}^{(t)}}\exp\left(\bF{\omega}_{\bF{u}}^\top \bF{m}_{\bF{x}}\right)\exp\left(-\delta_{\bF{u},k} (t-t_{\bF{x}})\right).\label{eqn:complete_model}
    \end{aligned}
\end{equation}
For notational brevity, we define $\Theta_{\bF{u}}:=(\gamma, \alpha, \bF{\beta}, \bF{\omega}, \delta_{k}, \delta_d, \delta_s)_{\bF{u}}$ as the set of parameters to be estimated for node $\bF{u}$ and $\Theta := \cup_{\bF{u}} \Theta_{\bF{u}}$ as the full set of parameters for $\stemmed$. These parameters can be estimated using the following} Maximum Likelihood Estimation (MLE) {problem}:
\begin{equation}\label{eq: MLE_full}
    \hat{\Theta}_{\text{MLE}}  
    = 
    \underset{\Theta}{\text{argmax }}  L(\Theta;\mc{D})
    ={}\underset{\Theta}{\text{argmax }}  \prod_{
    \bF{u} \in I\times S
    }
    \prod_{
        t \in \{t_{\bF{x}}: \bF{x}\in\mc{D}^{(T_{\bF{u}})}_{\bF{u}}\}
    } 
    \lambda_{\bF{u}}(t) \exp\Bigg(-\int_0^{T_{\bF{u}}} \lambda_{\bF{u}}(z) dz\Bigg),
\end{equation}
where ${L}$ represents the likelihood of STEMMED given history $\mc{D}$ that combines the history on every node $D_{\bF{u}}^{(T_{\bF{u}})}$ while $T_{\bF{u}}$ is the last observed event time at the node $\bF{u}$.

While solving \eqref{eq: MLE_full} requires estimating ~
{$|\Theta| = (p + q +5)^{|I\times S|}$}
parameters simultaneously, which may be undesirable both from computational and practical viewpoints, { we observe that \eqref{eq: MLE_full} can be decomposed into $|I\times S|$ subproblems, each of which corresponds to estimating parameters for a node, thereby enabling parameter estimation in parallel processes. Specifically, after applying logarithm transformation to \eqref{eq: MLE_full}, we can reorganize the resulting equation and obtain
\begin{equation} \label{eq: MLE_log_each}
    \widehat{\Theta}_{\bF{u}, \text{MLE}} 
    = \underset{\Theta_{\bF{u}}}{\text{argmax }} \ell_{\bF{u}}\left(\Theta_{\bF{u}} ; \mc{D}_{\bF{u}}^{(T_{\bF{u}})}\right) 
    = \underset{\Theta_{\bF{u}}}{\text{argmax }} \sum_{t \in \left\{t_{\bF{x}}: \bF{x}\in\mc{D}_{\bF{u}}^{(T_{\bF{u}})}\right\}} \log \lambda_{\bF{u}}(t) - 
    \int_{0}^{T_{\bF{u}}} \lambda_{\bF{u}}(z)dz.
\end{equation}}
The first term in equation \eqref{eq: MLE_log_each} relies on past event times and can be calculated by inserting these times into equation \eqref{eq: local_stemmed_final}. The second term is a definite integral of \eqref{eq: local_stemmed_final} dependent on data-generating processes for local information and event features. In our implementation, we assume these processes are left-continuous step functions with discontinuities at event times—an assumption commonly used in point process models with spatial covariates and individual features. This allows straightforward computation of the integral in \eqref{eq: MLE_log_each}. Further details are available in \S\ref{appendix: param_est_app} of our E-companion.
{
Importantly, decomposing \eqref{eq: MLE_full} to \eqref{eq: MLE_log_each} not only suggests an efficient node-by-node model-fitting procedure via parallel computing, but also promotes discussion of data sharing and cooperation framework between different communities or agencies, see the next section (\S\ref{sec: deployment}) for details. 
}

{\section{{Online Operation of STEMMED-Based Surveillance System}}
\label{sec: deployment}
}
In this section, we detail implementations and online operation for OOD surveillance using a STEMMED-based surveillance system. We first outline an online OOD forecast algorithm using STEMMED, then discuss operational adjustments including updating schedules and interagency cooperation between local public health departments. 


\subsection{OOD Forecast with STEMMED}\label{sec:online_framework}
    One of the most important tasks for STEMMED is to forecast the OOD progression in multiple time periods into the future to prepare for timely interventions, including medical, educational, and legal resource allocation. Generally, these forecasts from point process models can be performed using a simulation-based algorithm, i.e., the thinning algorithm \citep{ogata1981lewis}, whose main idea is to {simulate a non-stationary Poisson process by sampling candidate next-arrival times from an Exponential distribution and accept them based on the assigned CIF. Therefore, the events are generated one-by-one and are accounted for in the generation of the next events. See \cite{laub2015hawkes} and \cite{reinhart2018review} for further details.}
    {
    Since $\stemmed$ requires information beyond event times, such as spatial covariates and patient-level features, as shown in \eqref{eq: mu} -- \eqref{eq: g}, we develop Algorithm~\ref{algo: multi-step_pred} based on the the thinning algorithm. Specifically, for each simulated sample path, we generate the arrival time of the next event ($t_{\bF{x}}$) from an Exponential distribution whose parameter is the sum of intensities at current time $t$, i.e., $\sum_{\bF{u}}\lambda_{\bF{u}}(t)$. Then, we employ the thinning method. Specifically, this new event is accepted with probability $\frac{\sum_{\bF{u}}\lambda_{\bF{u}}(t_{\bF{x}})}{\sum_{\bF{u}}\lambda_{\bF{u}}(t)}$ and, if accepted, is assigned to some node $\bF{v}$ with probability $\frac{\lambda_{\bF{v}}(t_{\bF{x}})}{\sum_{\bF{u}}\lambda_{\bF{u}}(t_{\bF{x}})}$. Finally, we sample event features from the empirical distribution of features $\widehat{f}_{\bF{v}, \bF{m}_{\bF{x}}| \mc{D}_{\bF{v}}^{(t)}}$ on the assigned node $\bF{v}$ using data prior to time $t$. The sampled data are then added to the database to facilitate prediction for the next events. We note that the parameters of $\lambda_{\bF{u}}$ are not updated during the process, and, due to the simulation nature of the algorithm, each new event affects future intensities.
    }

{
\OneAndAHalfSpacedXI
\begin{algorithm}[t!]
\fontsize{8pt}{8pt} \selectfont

\caption{OOD Forecast with $\stemmed$}\label{algo: multi-step_pred}
\hspace*{\algorithmicindent} \textbf{Input}: $t, \mc{D}^{(t)}$, $\{\lambda_{\bF{u}}(t)\mid\bF{u}\in I\times S\}$, $T$\\
\hspace*{\algorithmicindent} \textbf{Output}: $\mc{D}^{(T)}$
\begin{algorithmic}[1]

\While{$t \leq T$}
    \State 
        $\lambda_{\text{max}} \gets \sum_{\bF{u}}\lambda_{\bF{u}}(t)$
        {\color{blue}\Comment{Compute $\lambda_{\bF{u}} (t)$ using \eqref{eqn:complete_model}}}
    \State 
        $t_{\bF{x}} \sim \text{Exponential}\left(\lambda_{\text{max}}\right)$        \Comment{Generate the next event time.}
    \State
        $\text{accept}\sim\text{Bernoulli}\left(
            \frac{\sum_{\bF{u}}{\lambda_{\bF{u}}(t_{\bF{x}}) }}{\lambda_{\text{max}}}
        \right)$
    \If{\text{accept} = 1}
        \State 
            $\bF{v}\sim\text{Categorical}\left(\frac{\lambda_{\bF{v}}\left(t_{\bF{x}}\right)}{\sum_{\bF{u}}{\lambda_{\bF{u}}(t_{\bF{x}}) }}\right)$
            \Comment{Assign event node.}
        \State
            $\bF{m}_{\bF{x}}\sim \widehat{f}_{\bF{v}, \bF{m}_{\bF{x}}| \mathcal{D}_{\bF{v}}^{(t)}}$
            \Comment{Generate feature from estimated feature distribution.}
        \State
            $\bF{x} \gets \left( t_{\bF{x}}, \bF{m}_{\bF{x}} \right)$
        \State 
            $\mc{D}^{(t)}_{\bF{v}} \gets \mc{D}^{(t)}_{\bF{v}}\bigcup\{\bF{x}\}$
        \State 
            $\mc{D}^{(t)} \gets \mc{D}^{(t)}\bigcup\mc{D}^{(t)}_{\bF{v}}$
        {\color{blue}\Comment{Update history for intensity computation.}}
    \EndIf
    \State $t \gets t_{\bF{x}}$
\EndWhile
\end{algorithmic}
\end{algorithm}
}

\subsection{Operational Adjustment of STEMMED to an OOD Surveillance System} \label{sec:stemmed-based_coop}
    { 
    In previous sections, we have developed and examined $\stemmed$ from a statistical modeling perspective. As our goal is to develop a reliable public health surveillance system, this section discusses several practical considerations for implementation and operation.
    
    First, Algorithm~\ref{algo: multi-step_pred} generates a single sample path according to $\stemmed$'s CIF, which is subject to multiple probabilistic realizations. Consequently, decisions made on a single forecast can be subject to great variability, which is undesirable for informing public health interventions. Moreover, while Algorithm~\ref{algo: multi-step_pred} is developed on continuous time, the forecasts can be more accessible if outcomes are collected and reported on a regular basis, e.g., weekly or monthly. To address these concerns, we generate multiple sample paths and collect each event into discrete timestamps using $\lceil t_{\bF{x}}\rceil$, then take sample averages across each sample to obtain a Monte Carlo-type mean estimate across timestamps. While our simulation-based approach is computationally more intensive than solving ODEs (as would be possible for standard Hawkes processes with exponential kernels), it readily enables computation of other distributional quantities beyond the mean, such as confidence intervals and tail probabilities for risk assessment, by analyzing the empirical distribution across multiple simulation runs. This approach not only reduces variability by smoothing out random fluctuations but also provides a more robust and interpretable basis when using $\stemmed$ for OOD surveillance.

    }

\section{Case Study: Opioid Overdose Deaths in Massachusetts}\label{sec: case_study}

{
In this section, we present a case study on the opioid crisis in Massachusetts (MA), USA, focusing on monthly illicit OODs at the city level to facilitate the generation of substantive managerial insights using STEMMED. We begin this section by introducing our dataset, pre-processing steps, and {setup} in \S \ref{sec:data_overview}. 
The objective of our case study is two-fold. First, we aim to characterize the dynamics of illicit opioid-related overdose deaths from historical data. Second, we aim to estimate the utility of STEMMED as a real-time surveillance system that can potentially inform public health interventions. Therefore, we divide our case study into two components: (i) offline analysis of STEMMED modeling (\S\ref{sec:offline_analysis}) and (ii) online real-time operation of a STEMMED-based surveillance system (\S\ref{sec:online_deploy}). Finally, we provide policy implications derived from our case study in \S\ref{sec:policy_implication}.

}

\subsection{Data Overview and Preprocessing}\label{sec:data_overview}

We compiled our dataset from three data sources: (1) the Massachusetts Registry of Vital Records and Statistics (MA-RVRS), (2) the U.S. Census Bureau, and (3) the Substance Abuse and Mental Health Services Administration (SAMHSA).

From MA-RVRS, we obtained individual-level death records ranging from January 1, 2015 to April 30, 2022. We used International Classification of Diseases, Tenth Revision, Clinical Modification (ICD-$10$-CM) codes to identify decedents under the three main underlying causes of death related to illicit opioid overdose \citep{ahmad2022nchs}, i.e., heroin, fentanyl and its derivatives, and psychostimulants. In addition to the underlying causes of death, we obtained a subset of information appearing on decedents' death certificates, including location and time of death and other individual-level socioeconomic information such as age, gender, and race. Over the study period, $15,584$ illicit OODs were recorded, with heroin, fentanyl, and psychostimulants accounting for $3,161$, $13,024$, and $8,638$ cases, respectively. More than half of these OODs involved more than one drug. Specifically, $47.37\%$ of OODs were attributed to a single drug type, $45.63\%$ to two drug types, and $7.05\%$ to all three drug types. Among all cities in MA, Boston represents the largest share at $14.26\%$ of OODs, followed by Worcester at $6.31\%$. Moreover, from the U.S. Census Bureau, we obtained city-level annual census reports ranging from $2014$ to $2022$. These data contain local socioeconomic features such as population, poverty level, and others. Finally, from SAMHSA we extracted the opioid treatment programs directory, which lists the name, address, and certification received date of the opioid treatment programs in MA. 

{The primary goal of this case study is to derive policy insights from the estimated $\stemmed$ model and to analyze its real-time online operations during this data period. For the purposes of our analysis, we have divided the time frame into monthly intervals, thereby advancing $t$ by $1$ implies $1$ month progression.} For the local communities, we considered the top $25$ cities in Massachusetts ($|I| = 25$) where most OODs occurred, namely, Boston, Worcester, Springfield, Brockton, New Bedford, Lowell, Fall River, Lawrence, Lynn, Weymouth, Quincy, Salem, Everett, Barnstable, Taunton, Haverhill, Cambridge, Pittsfield, Leominster, Holyoke, Attleboro, Plymouth, Framingham, Wareham, and Melrose. {The spatial unit for distance between cities is miles, for which we calculated using Vincenty's formula based on the coordinates of city centers.} For OOD types, we considered three illicit opioid-related causes of death ($|S| = 3$), namely, heroin, fentanyl, and psychostimulants. { This spatio-temporal framework, with monthly intervals across city-OOD type nodes, aligns with public health management practices, where such reporting enables detailed tracking of OOD patterns and informs targeted interventions.} As for the features, we incorporated the ones not directly obtainable from death certificates for the \textit{nodal features} {($\bF{y}_{\bF{u},t}$)}. These features are: population {(\texttt{POP})}, poverty rate {(\texttt{POV})}, and treatment program coverage {(\texttt{TPC})}. {While \texttt{POP} refers to the total residents in city $i$, \texttt{POV} represents the percentage of residents in city $i$ living below the federal poverty line and \texttt{TPC} is the serving population of a treatment program.} For \textit{individual-level features} {($\bF{m}_{\bF{x}}$)}, we selected age, sex, race and ethnicity, and poly-substance involvement, which have been shown to be highly associated with OODs \citep{scholl2019drug}. We applied one-hot encoding for sex, race, and ethnicity. The resulting features are: Age {(\texttt{AGE})}, Male {(\texttt{SEX}),} White, non-Hispanic {(\texttt{WNH})}, Black, non-Hispanic {(\texttt{BNH})}, Hispanic {(\texttt{HISP})}, and Polysubstance Involvement {(\texttt{PLY})}. We calculated social connectivity {($\bF{\theta}_{\bF{u},t}^{\bF{v}}$)} by the percentage of OOD
events at community $i$ and $j$ that involved both drugs $s$ and $s'$, prior to time $t$. Finally, we standardized \texttt{POP}, \texttt{POV}, \texttt{TPC}, and \texttt{AGE} to stabilize parameter estimation. We refer readers to E-companion \ref{sec:add_detaild_case_study} for additional details.


{
\subsection{Offline Analysis}\label{sec:offline_analysis} 
{
For offline analysis, we apply $\stemmed$ to the full dataset to analyze policy insights gained from $\stemmed$'s unique model structure. To this end, we investigate the contribution of the features and parameters and explore distinct perspectives on the dynamics of OOD, from the view of substance and spatial classifications.
}

{
\SingleSpacedXI
\begin{table}[htb]

\centering
\caption{
Clustered Mean Estimates (Standard Deviations) of $\stemmed$ Parameters by OOD Types and Counties. Values are Scaled by $\mathbf{10^{-2}}$. Largest Values for Nodal, Individual, and Structural Parameters for each Subgroup are Highlighted. (Abbreviations: \texttt{POP} = Population, \texttt{POV} = Poverty Rate, \texttt{TPC} = Treatment Program Coverage, \texttt{PLY} = Polysubstance Involvement, \texttt{AGE} = Age, \texttt{SEX} = Male, \texttt{WNH} = White, non-Hispanic, \texttt{BNH} = Black, non-Hispanic, \texttt{HISP} = Hispanic)
}
\resizebox{1.01\textwidth}{!}{%
\begin{tabular}{|c|lll|llllllllll|}
\hline
                                                       & \multicolumn{3}{c|}{OOD Types}                                                                   & \multicolumn{10}{c|}{Counties}                                                                                                                                                                                                                                                                                     \\
                                                        Parameters& \multicolumn{1}{c}{Heroin} & \multicolumn{1}{c}{Fentanyl} & \multicolumn{1}{c|}{Psycho-} & \multicolumn{1}{c}{Bristol} & \multicolumn{1}{c}{Middlesex} & \multicolumn{1}{c}{Suffolk} & \multicolumn{1}{c}{Worcester} & \multicolumn{1}{c}{Barnstable} & \multicolumn{1}{c}{Essex} & \multicolumn{1}{c}{Hampden} & \multicolumn{1}{c}{Norfolk} & \multicolumn{1}{c}{Plymouth} & \multicolumn{1}{c|}{Berkshire} \\ 
                                                        & \multicolumn{1}{c}{} & \multicolumn{1}{c}{} & \multicolumn{1}{c|}{stimulant} & \multicolumn{1}{c}{} & \multicolumn{1}{c}{} & \multicolumn{1}{c}{} & \multicolumn{1}{c}{} & \multicolumn{1}{c}{} & \multicolumn{1}{c}{} & \multicolumn{1}{c}{} & \multicolumn{1}{c}{} & \multicolumn{1}{c}{} & \multicolumn{1}{c|}{} \\ \hline
                                                       \textbf{Nodal}& \multicolumn{3}{c|}{}& \multicolumn{10}{c|}{}   \\
\multicolumn{1}{|l|}{$(\bF{\beta})$}      & \multicolumn{1}{c}{}       & \multicolumn{1}{c}{}         & \multicolumn{1}{c|}{}                & \multicolumn{1}{c}{}        & \multicolumn{1}{c}{}          & \multicolumn{1}{c}{}        & \multicolumn{1}{c}{}          & \multicolumn{1}{c}{}           & \multicolumn{1}{c}{}      & \multicolumn{1}{c}{}        & \multicolumn{1}{c}{}        & \multicolumn{1}{c}{}         & \multicolumn{1}{c|}{}          \\
\texttt{POP}                                                     & 0.6 (0.5)              & {0.9} (0.7)                & 0.8 (0.6)                        & 0.8 (0.7)               & 0.4 (0.4)                 & \textbf{1.6} (0.3)               & 0.9 (0.6)                 & 0.8 (0.5)                  & 0.7 (0.4)             & 1 (0.6)               & 1.2 (0.5)               & 0.6 (0.5)                & 1.2 (0.9)                 \\
\texttt{POV}                                                     & \textbf{0.9} (0.5)              & \textbf{2.8} (1.8)                & \textbf{2.1} (1.3)                        & \textbf{1.7} (1.1)               & \textbf{2.4} (1.9)                 & \textbf{1.6} (1.3)               & \textbf{1.6} (1.2)                 & \textbf{1.6} (1.4)                  & \textbf{1.7} (1.8)             & \textbf{1.8} (1)               & \textbf{2.5} (1.8)               & \textbf{1.9} (1.6)                & \textbf{1.7} (0.8)                 \\
\texttt{TPC}                                                     & 0.4 (0.3)              & 1 (0.6)                & {1.2} (0.6)                        & 0.9 (0.5)               & 1 (0.6)                 & 0.7 (1)               & 1.1 (0.8)                 & {1.3} (0.6)                  & 0.9 (0.6)             & 0.4 (0.4)               & 0.6 (0.4)               & 1 (0.6)                & 0.7 (0.8)                 \\
                                                       \textbf{Individual}& \multicolumn{3}{c|}{}                                                                   & \multicolumn{10}{c|}{}   
\\
\multicolumn{1}{|l|}{$(\bF{\omega})$} & \multicolumn{1}{c}{}       & \multicolumn{1}{c}{}         & \multicolumn{1}{c|}{}                & \multicolumn{1}{c}{}        & \multicolumn{1}{c}{}          & \multicolumn{1}{c}{}        & \multicolumn{1}{c}{}          & \multicolumn{1}{c}{}           & \multicolumn{1}{c}{}      & \multicolumn{1}{c}{}        & \multicolumn{1}{c}{}        & \multicolumn{1}{c}{}         & \multicolumn{1}{c|}{}          \\
$\texttt{PLY}$                                          & 4.4 (2.2)                  & 5.1 (2.3)                    & 5.7 (3.8)                            & 5.3 (3.1)                   & 5.5 (3.0)                     & 4.7 (4.2)                   & 4.9 (1.9)                     & 2.6 (1.7)                      & 5.7 (2.7)                 & 6.2 (2.4)                   & 3.6 (3.6)                   & 4.9 (3.0)                    & 4.2 (3.6)                     \\
$\texttt{AGE}$                                          & \textbf{8.8} (4.9)                  & \textbf{8.3} (4.1)                    & 8.5 (4.4)                            & \textbf{9.8} (4.7)                   & \textbf{8.0 }(3.9)                     & 5.5 (5.9)                   & 8.8 (5.5)                     & 6.3 (5.9)                      & 8.2 (4.4)                 & 6.7 (4.1)                   & \textbf{10.7} (4.3)                  & \textbf{10.5} (3.4)                   & 6.3 (5.2)                     \\
$\texttt{SEX}$                                          & 7.1 (4.6)                  & 7.9 (5.4)                    & \textbf{10.4} (5.2)                           & 7.4 (5.1)                   & 7.4 (5.8)                     & \textbf{11.2} (3.8)                  & 7.3 (5.9)                     & 7.1 (8.0)                      & \textbf{10.4} (4.2)                & \textbf{10.9} (3.6)                  & 10.6 (5.0)                  & 8.7 (5.3)                    & 1.7 (1.4)                     \\
$\texttt{WNH}$                                          & 6.3 (4.7)                  & 6.8 (3.5)                    & 6.9 (3.7)                            & 7.6 (3.4)                   & 6.6 (4.1)                     & 4.2 (3.4)                   & \textbf{7.8} (3.6)                     & 6.2 (2.1)                      & 5.6 (3.7)                 & 4.6 (3.7)                   & 8.9 (5.2)                   & 4.9 (3.1)                    & \textbf{13.0} (1.8)                    \\
$\texttt{BNH}$                                          & 8.5 (3.8)                  & 7.6 (4.1)                    & 5.8 (3.9)                            & 7.9 (3.8)                   & 7.1 (4.1)                     & 6.6 (2.6)                   & 7.6 (3.9)                     & \textbf{8.9} (5.7)                      & 5.5 (4.5)                 & 8.2 (2.8)                   & 6.5 (3.4)                   & 7.5 (4.5)                    & 10.3 (7.1)                    \\
$\texttt{HISP}$                                         & 6.0 (3.3)                  & 4.7 (2.8)                    & 6.7 (3.9)                            & 6.8 (3.1)                   & 4.6 (2.0)                     & 7.5 (6.0)                   & 5.5 (3.3)                     & 8.0 (1.3)                      & 6.3 (3.9)                 & 5.1 (3.8)                   & 6.4 (4.6)                   & 4.6 (3.4)                    & 6.6 (4.5)                     \\
\multicolumn{1}{|l|}{{\ul \textit{\textbf{Structural}}}}     & \multicolumn{1}{c}{}       & \multicolumn{1}{c}{}         & \multicolumn{1}{c|}{}                & \multicolumn{1}{c}{}        & \multicolumn{1}{c}{}          & \multicolumn{1}{c}{}        & \multicolumn{1}{c}{}          & \multicolumn{1}{c}{}           & \multicolumn{1}{c}{}      & \multicolumn{1}{c}{}        & \multicolumn{1}{c}{}        & \multicolumn{1}{c}{}         & \multicolumn{1}{c|}{}          \\
$\delta_k$                                                & 398 (130)              & 219 (36)                 & 295 (45)                         & 332 (112)               & 292 (143)                 & 256 (57)                & 287 (118)                 & \textbf{355} (127)                  & \textbf{318.4} (97)              & 267 (38)                & 254 (135)               & \textbf{344} (103)                & 277 (26)                  \\
$\delta_d$                                               & \textbf{444} (45)               & 267 (11)                 & 257 (69)                         & 325 (80)                & \textbf{305} (99)                  & 341 (130)               & 334 (86)                  & 323 (77)                   & 304 (124)             & 353 (109)               & 324.3 (107)               & 330.4 (105)                & \textbf{354} (125)                 \\
$\delta_s$                                               & 275 (199)              & \textbf{312} (156)                & \textbf{380} (114)                        & \textbf{336} (176)               & 269 (171)                 & \textbf{452} (74)                & \textbf{393} (163)                 & 265 (191)                  & 308 (164)             & \textbf{393} (46)                & \textbf{377} (182)               & 290 (198)                & 225 (61)                  \\
\hline
\end{tabular}%
}
\label{table:estimated_params}
\end{table}
}

\paragraph{Estimated Parameters.} 
{
{With $75$ sets of estimated parameters, each corresponding to a city-OOD node, we grouped the results by $3$ OOD types, i.e., heroin, fentanyl, and psychostimulants, and by $10$ counties, i.e., Bristol, Middlesex, Suffolk, Worcester, Barnstable, Essex, Hampden, Norfolk, Plymouth, and Berkshire, to provide analyses of overall trends from substance- and location-centered perspectives. This approach allows us to explore variations in parameter estimates across OOD categories and geographic regions, which facilitates identification of distinct trends in any subgroups.

Table \ref{table:estimated_params} shows the mean estimates and standard deviations (in parentheses) of STEMMED parameters across nodes within each category. Poverty rate consistently has the strongest influence on the background event rate across all OOD types, with mean estimates of $0.009$ for heroin-related, $0.028$ for fentanyl-related, and $0.021$ for psychostimulant-related OODs. For heroin-related OODs, population and treatment program coverage follow in influence. In contrast, fentanyl and psychostimulant OODs show greater effects from treatment program coverage ($0.010$ and $0.012$, respectively) than population ($0.009$ and $0.008$, respectively). Notably, while population is the second most influential factor for heroin-related OODs, it has the smallest impact for fentanyl- and psychostimulant-related OODs, highlighting substance-specific differences in socio-demographic effects. At the county level, poverty rate remains the dominant factor, with the largest mean estimate in Norfolk ($0.025$) and the smallest in Worcester, Suffolk, and Barnstable (each $0.016$). Treatment program coverage is most influential in Barnstable ($0.01$) and least in Hampden ($0.004$), while population has the greatest effect in Suffolk ($0.016$). These results reveal both substance-specific and geographic variations in OOD risk factors, emphasizing the need for tailored interventions.

Next, we examine the estimated parameters for individual features contributing to triggering effects to analyze substantial heterogeneity across OOD types. When grouped by OOD type, age is the most influential factor for heroin- ($0.088$) and fentanyl-related OODs ($0.083$), while sex has the strongest association with psychostimulant-related OODs ($0.104$), followed by age ($0.085$). Polysubstance involvement has the smallest effect on the heroin- ($0.044$) and psychostimulant-related OODs ($0.057$), whereas Hispanic ethnicity has the least association on fentanyl-related OODs ($0.047$). While heroin-related OODs show relatively consistent feature effects, fentanyl and psychostimulant OODs display greater heterogeneity. For instance, the effect of sex on psychostimulant OODs is nearly double that of polysubstance involvement. County-cluster analysis further reveals geographic differences in triggering dynamics. Age is the most influential factor in Bristol ($0.098$), Middlesex ($0.080$), Worcester ($0.088$), Norfolk ($0.107$), and Plymouth ($0.105$), suggesting older individuals in these areas have higher association with OOD cases. In contrast, sex is most impactful in urban counties like Suffolk ($0.112$), Essex ($0.104$), and Hampden ($0.109$), indicating a higher likelihood of OOD cases among males. Notably, Black, non-Hispanic OODs have the strongest influence in Barnstable ($0.089$), while White, non-Hispanic OODs dominate in Berkshire ($0.130$). 

Finally, in $\stemmed$, parameters $\delta_k$, $\delta_d$, and $\delta_s$ shape OOD spatiotemporal dynamics, with $\delta_k$ and $\delta_d$ control temporal and spatial decay, while $\delta_s$ boosts social connectivity effects. Table \ref{table:estimated_params} shows heroin’s associated coefficient $\delta_k$ ($3.98$) yields the fastest temporal decay (halving in around 5 days), compared to 9.5 (fentanyl) and 7 (psychostimulants) days. For spatial decay ($\delta_d$), heroin’s coefficient ($4.44$) exceeds fentanyl ($2.67$) and psychostimulants ($2.578$). On the other hand, the social connectivity parameter $\delta_s$ is highest for psychostimulants ($3.80$), followed by fentanyl ($3.12$) and heroin ($2.75$). These results suggest that fentanyl-related OODs are influenced by past events for a longer duration, compared to the other OOD types. Moreover, while heroin OODs are more localized and fentanyl's effects spread farther spatially, psychostimulants are most sensitive to shared drug-use behaviors, which may drive stronger inter-node spillovers. Clustering by counties further reveals geographic variations in spatiotemporal dynamics. Barnstable has the highest mean estimated coefficient $\delta_k$ ($3.55$), indicating that OOD triggering effects generally last more than 5 days across all counties. For spatial decay ($\delta_d$), Berkshire sho ws the fastest rate ($3.54$), followed by Hampden ($3.53$) and Suffolk ($3.41$), while Middlesex has the slowest ($3.05$). Regarding social connectivity ($\delta_s$), Suffolk exhibits the highest value ($4.52$), followed by Hampden ($3.93$) and Worcester ($3.93$), with Berkshire showing the lowest ($2.25$). These patterns suggest that OOD cases in Barnstable dissipates quickly, which requires rapid response, while Norfolk experiences more persistent effects and can benefit from longer-term interventions. Spatial spillovers are most localized in Berkshire, while Middlesex shows broader influence, possibly due to its central location. Additionally, urban counties like Suffolk exhibit stronger polysubstance spillovers, where shared drug-use behaviors can largely drive OODs.}
}
\paragraph{Dynamic Network Connections.} 
{

We next explore the spatiotemporal dynamics of OOD connections in MA by calculating the mean inter-drug and inter-city interactions across three timestamps, i.e., January 2016, January 2019, and January 2022, and scaling the results by their maximum magnitudes for comparison. Specifically, we define the average inter-drug interactions as 
$ A_{i,j}(t) := \frac{\sum_{s \in S} \sum_{s' \in S} A_{\bF{u}}^{\bF{v}}(t)}{2|S|} $
and the average inter-city interactions as $ A^{s,s'}(t) := \frac{\sum_{i \in I} \sum_{j \in I} A_{\bF{u}}^{\bF{v}}(t)}{2|I|}$. To provide comparisons within and across timestamps, we normalize these interactions by $ \hat{A}_{i,j}(t) := \frac{A_{i,j}(t)}{\max_t A_{i,j}(t)} $ and $ \hat{A}^{s,s'}(t) := \frac{A^{s,s'}(t)} {\max_t A^{s,s'}(t)} $. The results are presented in Figure~\ref{fig:dynamic_connections}, with Figure~\ref{fig:ood_to_ood_conn_3yr} illustrating the average inter-drug-type connections and Figure~\ref{fig:city_to_city_conn_3yr} displaying the top $5\%$ average strength of connections between different cities on MA county map.
}
{
\SingleSpacedXI
\begin{figure}[thb]
    \centering
    \begin{minipage}{0.9\textwidth}
    \centering
    \subfloat[
    {Temporal Evolution of OOD-to-OOD Connections.} ]{\includegraphics[width=0.95\textwidth,keepaspectratio]{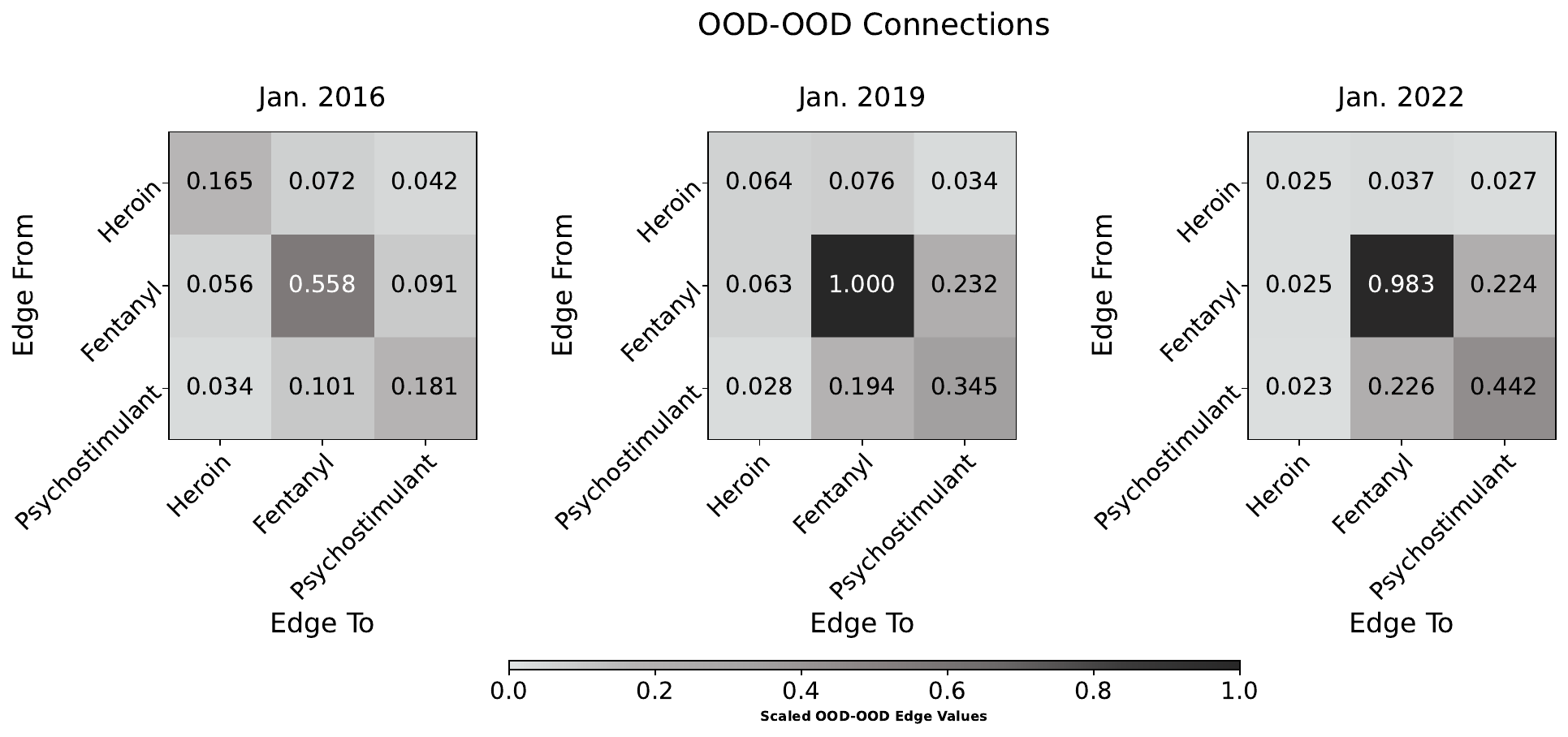}
    \label{fig:ood_to_ood_conn_3yr}}
    
    \subfloat[{Temporal Evolution of City-to-City Connections on Massachusetts County Map}.]{\includegraphics[width=0.95\textwidth,keepaspectratio]{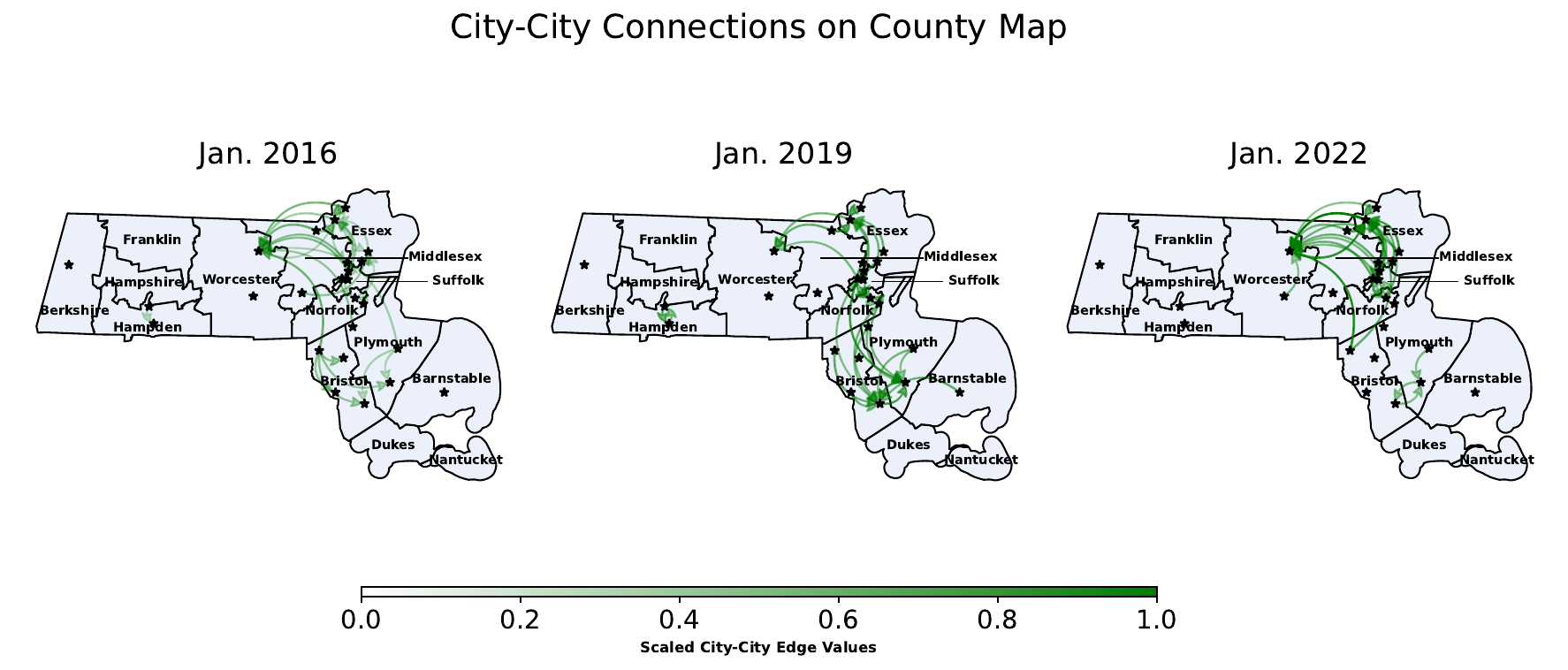}
    \label{fig:city_to_city_conn_3yr}}
    
    \caption{
    { Spatiotemporal Dynamics of OOD Connections: Drug-Type and City Interactions in Massachusetts. (a) OOD-to-OOD connections over time, and (b) city-to-city connections on a county map.}
    }
    \label{fig:dynamic_connections}
\end{minipage}
\end{figure}
}

{

From Figure \ref{fig:ood_to_ood_conn_3yr}, we observe distinct patterns of OOD across drug types. Within-drug type connections demonstrate differential temporal dynamics. Notably, fentanyl-to-fentanyl connections exhibit the most pronounced growth, escalating from a normalized strength of $0.558$ in January 2016 to $1.0$ in January 2019, with a marginal decline to $0.983$ by January 2022. Conversely, psychostimulant-to-psychostimulant connections show a steady increase from $0.181$ to $0.442$ over the same period, while heroin-to-heroin connections substantially decrease from $0.165$ to $0.025$.
Cross-drug type interactions further reveal interdependencies between OOD types. The fentanyl-to-psychostimulant connection increases from $0.091$ in January 2016 to $0.232$ in January 2019, before slightly receding to $0.224$ by January 2022. Simultaneously, the psychostimulant-to-fentanyl connection rises from $0.101$ to $0.194$ in January 2019 and further to $0.226$ in January 2022. These trends align with the high social connectivity parameters for fentanyl and psychostimulants observed in previous analyses, which implies evolving shared drug-use behaviors during study period.

The between-city connections illustrated in Figure \ref{fig:city_to_city_conn_3yr} unveil distinct regional OOD dynamics. Central and western counties, including Berkshire and Hampden, demonstrate localized connection patterns wherein city-to-city connection magnitudes remain consistently low, typically below $0.2$, with minimal fluctuations across the study period. This localization corresponds to larger spatial decay parameters, indicating geographic constraints potentially stemming from lower population density and limited inter-city mobility.
Southeastern counties (Bristol, Barnstable, and Plymouth) present a more dynamic connectivity profile. Connection strengths generally oscillate between $0.2$ in January 2016 and $0.8$ by January 2019, subsequently dropping below $0.2$ by January 2022. Notably, these cities exhibit transient interconnections with northern cities from 2016 to 2019, before becoming largely isolated by 2022.
The northeastern region, including Essex, Middlesex, Suffolk, and Norfolk counties, emerges as a critical OOD connection hub. Connection strengths range from $0.6$ to $0.8$ in January 2016, progressively increasing to $0.8$–$1.0$ by January 2019 and maintaining this level through January 2022. This region demonstrates a particularly noteworthy trend of increasing interconnectivity, with cities like Lowell and Lawrence becoming integrated into a tightly connected network centered around Boston.
The sustained strength of northeastern connections reflects the region's significant role in OOD activity. The high connectivity likely stems from dense population centers and complicated social networks. Ultimately, these findings suggest that urban proximity and shared drug markets may facilitate rapid OOD event transmission across interconnected urban landscapes.
}

{
\SingleSpacedXI
\begin{figure}[thb]
    \centering
    \begin{minipage}{0.9\textwidth}
    \centering
    \includegraphics[width=0.95\textwidth,keepaspectratio]{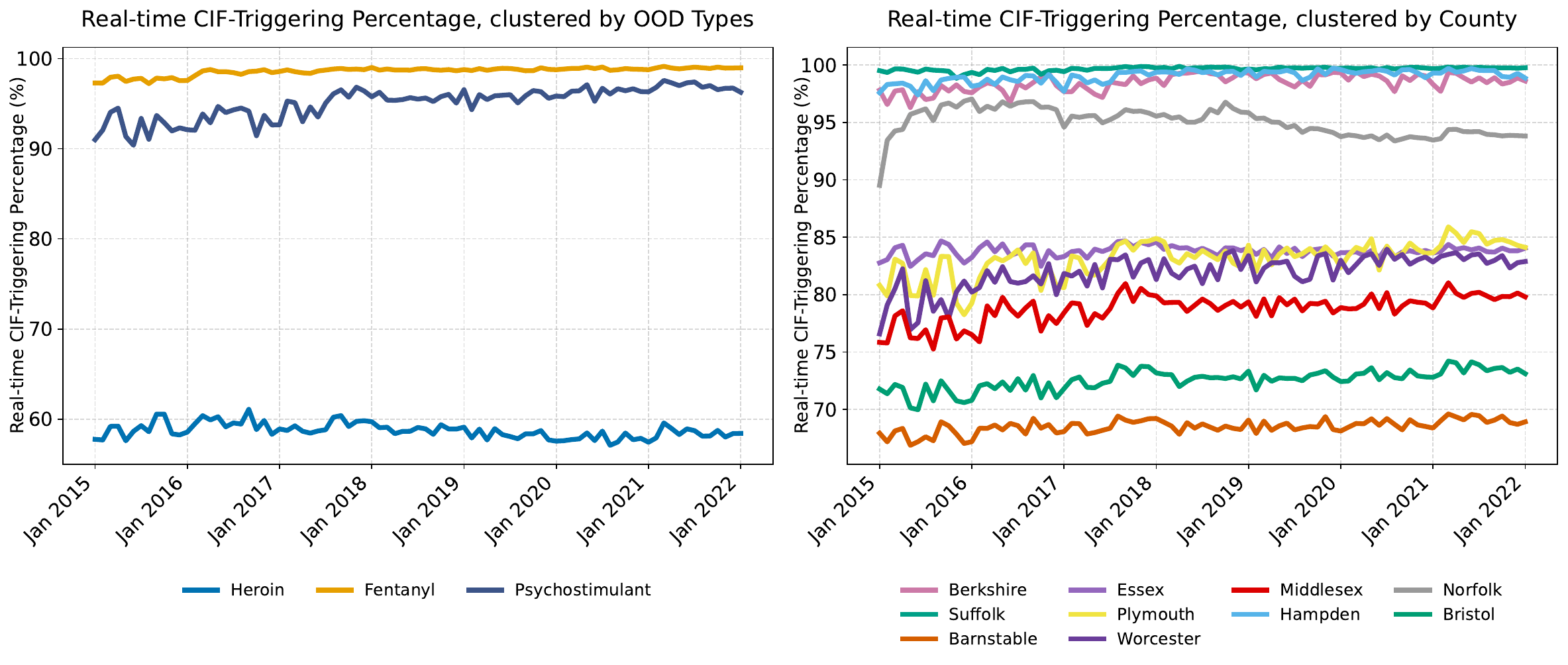}
    \caption{
Real-time CIF-Triggering Percentage Observed in Each City Over the Study Period.
    }
    \label{fig:trigger_perc}
\end{minipage}
\end{figure}
}

{
\paragraph{Dynamic Triggering Effects.} 
In Figure \ref{fig:trigger_perc}, we visualize the Real-time CIF-Triggering Percentage, which represents the ratio of  triggering effects in the CIF function at sample-path level, averaged across OOD types or counties. Specifically, for each timestamp, we calculate the CIF-trigger percentage as
$f_{\bF{u}}(t) := (1-\frac{\mu_{\bF{u}}(t)}{{\lambda_{\bF{u} (t) }}})\times100\%$. Next, we compute the average triggering percentage for drug type $i$ as
$ \hat{f}_{i}(t) := \frac{\sum_
{s\in S} f_{\bF{u}}(t)}{|S|}$ and for county $C$ as
$ \hat{f}^{{C}}(t) := \frac{\sum_
{i\in I}\sum_{s\in{C}} f_{\bF{u}}(t)}{|I\cup{C}|}$. Notably, this interpretation offers insight into the contribution of individual OOD events to the intensities. However, it does not establish causation and may be affected by model misspecification, as pointed out by \cite{reinhart2018self}.

From Figure \ref{fig:trigger_perc}, we observe distinct patterns in the average real-time CIF-triggering percentages of heroin, fentanyl, and psychostimulant-related OODs. Fentanyl demonstrates a consistently high value, exceeding $95\%$ throughout the study period. Psychostimulant OODs similarly show an increasing trend, rising from approximately $90\%$ to $98\%$. These results suggest that fentanyl and psychostimulant overdoses have a heightened capability to trigger subsequent incidents, potentially indicating a self-reinforcing cycle driven by their potency and pervasive presence in the drug supply.
In contrast, heroin-related OODs exhibit a more variable triggering pattern, fluctuating between $55\%$ and $62\%$. A notable declining trend emerges after 2019, which may reflect the ongoing transformation of the opioid crisis, with synthetic opioids like fentanyl increasingly making up a larger proportion of illicit drug use across the country.

From the county clustering perspective, Figure \ref{fig:trigger_perc} highlights significant variability of OOD triggering behavior across counties in MA. Suffolk County consistently demonstrates the highest real-time CIF-triggering percentages, consistently exceeding $97\%$ throughout the study period. This pattern likely stems from Suffolk's urban environment, where dense populations and established drug distribution networks facilitate rapid overdose event propagation. Berkshire and Hampden also display high and rising values, consistently above $95\%$. While further investigation is warranted, this observation suggests potential common underlying factors driving OOD events in these less urban areas. Conversely, Bristol and Barnstable exhibit the lowest real-time CIF-triggering percentages across the study period. This result suggest these areas have more background-type events; however, these values consistently remain above $65\%$, which highlights the critical importance of comprehensive investigation and interconnected OOD case analysis.
}

{
\subsection{Real-Time Operation of the STEMMED-Based Forecast System}\label{sec:online_deploy}
}
{
Our goal is to develop a reliable public health surveillance system; therefore, in this section we examine performance of the developed $\stemmed$ model and several established models, under practical considerations.
}
{
\paragraph{Real-Time Illicit OODs Forecasting.} 
To investigate the benefit of incorporating individual-level data, city-level features, and the network structure for forecasting trends, we included standard time series models, e.g., the Autoregressive Integrated Moving Average model (ARIMA) and the Vector Auto-Regressive model (VAR), point process models, e.g., SEPP and MEPP, and neural network, e.g., Graph Neural Network (GNN), into our study. ARIMA and VAR are commonly employed time series models to capture temporal dependencies in aggregated data and forecast future trends while GNN incorporates sophisticated spatio-temporal information into their model frameworks. Like MEPPs, the VAR, and GNN models can incorporate network connections between event streams. {Additionally, we consider forecasting using only the background rate learned by $\stemmed$, i.e., $\stemmed_{\text{bg}}$, to verify the advantages of including triggering effects. Note that $\stemmed_{\text{bg}}$ reduces to a non-homogeneous Poisson process with rate ${\mu}_{\bF{u}} (t)$ for node $\bF{u}$. We also consider $\stemmed$ with a different triggering function, i.e., the power-law kernel $f_{\delta}(t-t_{\bF{x}}) = (1/t-t_{\bF{x}})^{1+\delta}$ for both temporal and spatial parameters $\delta_k$ and $\delta_d$. This kernel generally gives slower decay rate compared to the exponential kernel used in the current setting (See \eqref{eq:stemmed_kappa}), which facilitates analyses of how the effects of previous OOD events diminish. We repeat Algorithm~\ref{algo: multi-step_pred} $100$ times and map the samples into monthly predictions and average across sample paths. Further details regarding model specifications are provided in E-Companion \ref{sec:spec_model}. 

Moreover, as described in \S\ref{sec:stemmed-based_coop}, we aim to investigate the impact of various data-sharing policies on $\stemmed$. We grouped nodes according to cluster assignments, with the constraint that OOD data and forecasts from node $\bF{v}$ are only accessible to node $\bF{u}$ if they belong to the same cluster. Our experimental design consists of three distinct policies: Policy $\pi_1$ reflects the practice of selective reporting wherein local communities share data exclusively for a specific OOD type. In other words, data-sharing exists only for $\{\bF{u} = (i,s)\in I\times S | i = j \}$ for OOD type $i$ and $j$. For example, local communities in $\stemmed$'s network only shares OOD information related to heroin, fentanyl, or psychostimulants. For Policy $\pi_2$, we assume data sharing is limited to communities within the same geographical class, such as counties, i.e., $\{\bF{u} = (i,s)\in I\times S | s \in C \}$ for county $C$. This policy reflects existing regional affiliations in public health actions. Finally, Policy $\pi_3$ restricts data exchange to neighboring communities in the same geographical clusters on the map, e.g., clusters obtained by using coordinates, i.e., $\{\bF{u} = (i,s)\in I\times S | s \in C' \}$ for some cluster $C'$. In our analyses, $\pi_3$ is split to $\pi_{3,3}$ with three coordinate-based clusters and $\pi_{3,5}$ with five clusters. We note that the $\stemmed$-based system that all $75$ nodes share their information with each other can be viewed as policy $\pi_3$ with $1$ cluster.
}

{ To incorporate other practical considerations in model operation}, we assume the database update information on a monthly basis in this model-based surveillance scenario, accompanied by model updates and a six-month ahead forecasts. 
{
We then calculate the mean absolute relative error (MARE) with a rolling window approach. Specifically, the absolute relative error is defined as $\text{ARE}_{\bF{u}}(t) = \frac{|\hat{Y}_{\bF{u}, t}-Y_{\bF{u}, t}|}{Y_{\bF{u}, t}+1}$ with $Y_{\bF{u}, t}$ ($\hat{Y}_{\bF{u}, t}$) representing OOD count (predictions) on node $\bF{u}$ at $t$-month ahead when the forecasts are made. We add $1$ to the denominator to address data scarcity issues, i.e., no OOD incidence at some month.
Then,
$\text{ARE}_{\bF{u}}(t)$
}is aggregated and averaged over the study period { and the nodes} to construct a rolling window prediction error for forecasts one to six months ahead. {Overall, MARE facilitates interpretation of models' capability in making long-term forecasts. We initiate the first forecasts starting from the 12th month.}

{
{
\SingleSpacedXI
\begin{table}[htb]

\centering
\caption{
\fontsize{8pt}{8pt} \selectfont
     Rolling-Window Mean Absolute Relative Error (MARE) in Model-Based Online Forecasting. $\stemmed_{\text{bg}}$ and $\stemmed_{\text{pl}}$ represents $\stemmed$ implemented with background rate or power-law kernel and $\stemmed_{\bF{\pi_{\cdot}}}$ represents $\stemmed$ implemented with data-sharing policy $\bF{\pi_{\cdot}}$.
}
\scalebox{0.7}{%
\begin{tabular}{|c|llllll|}
\hline
\multicolumn{1}{|l|}{}   & \multicolumn{6}{c|}{Forecast Period ($n$-Month Ahead)}                                                                                                                       \\ 
Models                 & \multicolumn{1}{c}{$1$} & \multicolumn{1}{c}{$2$} & \multicolumn{1}{c}{$3$} & \multicolumn{1}{c}{$4$} & \multicolumn{1}{c}{$5$} & \multicolumn{1}{c|}{$6$} \\ \hline
ARIMA                  & 0.682                   & 0.686                   & 0.689                   & 0.687                   & 0.694                   & 0.696                   \\
VAR                    & 0.751                   & 0.686                   & 0.686                   & 0.69                    & 0.698                   & 0.754                   \\
GNN                    & 0.713                   & 0.769                   & 0.786                   & 0.842                   & 0.781                   & 0.77                    \\
SEPP                   & 0.52                    & 0.537                   & 0.557                   & 0.576                   & 0.594                   & 0.626                   \\
MEPP                   & 0.536                   & 0.544                   & 0.562                   & 0.597                   & 0.613                   & 0.656                   \\
$\stemmed$             & \textbf{0.446}          & \textbf{0.457}          & \textbf{0.471}          & \textbf{0.485}          & \textbf{0.499}          & \textbf{0.516}          \\
$\stemmed_{\text{bg}}$             &0.642	&0.642	&0.639	&0.639	&0.636	&0.635          \\
$\stemmed_{\text{pl}}$ & 0.518                   & 0.524                   & 0.531                   & 0.548                   & 0.566                   & 0.594                   \\
$\stemmed_{\pi_1}$     & 0.467                   & 0.49                    & 0.501                   & 0.51                    & 0.522                   & 0.532                   \\
$\stemmed_{\pi_2}$     & 0.521                   & 0.538                   & 0.554                   & 0.573                   & 0.59                    & 0.554                   \\
$\stemmed_{\pi_{3,3}}$ & 0.474                   & 0.494                   & 0.502                   & 0.512                   & 0.518                   & 0.529                   \\
$\stemmed_{\pi_{3,5}}$ & 0.527                   & 0.538                   & 0.552                   & 0.565                   & 0.578                   & 0.593    \\ \hline        
\end{tabular}%
}
\label{table:forecast_error}
\end{table}
}
}

{
The results in Table~\ref{table:forecast_error} demonstrate $\stemmed$'s superior performance across all forecast horizons. For 1-month-ahead predictions, $\stemmed$ achieves the lowest Mean Absolute Relative Error (MARE) of 0.446, significantly outperforming ARIMA (0.682), VAR (0.751), GNN (0.713), SEPP (0.52), and MEPP (0.536). This performance advantage persists for longer horizons, with $\stemmed$'s MARE increasing to 0.516 at six months, compared to 0.696 for ARIMA, 0.754 for VAR, 0.770 for GNN, 0.626 for SEPP, and 0.656 for MEPP. For time-series models, ARIMA shows consistently high errors across all horizons, which reflects its inability to model spatial and network-driven OOD event dynamics. VAR, despite incorporating network connections, performs poorly, which suggests challenges in capturing complex OOD data interactions and temporal dependencies, especially in short-term ($1$-month) and longer-terms ($5$ and $5$ moths) predictions. GNN exhibits significant instability, with MARE fluctuating between 0.713 and 0.842, which may due to overfitting in modeling OOD trends. As for point process models, $\stemmed_{\text{bg}}$ outperforms ARIMA, VAR, and GNN but remains generally less effective than other point process counterparts, which highlights the importance of capturing triggering effects in forecasting. Notably, the strong performance of SEPP (0.52 MARE) and MEPP (0.536 MARE) for 1-month-ahead forecasts demonstrates that self- and mutual-excitation are fundamental features of OOD dynamics. However, their errors increase more rapidly over time, reaching 0.626 and 0.656 at 6 months, respectively. This degradation, contrasted with $\stemmed$'s sustained performance, suggests that time-varying parameters and individual-level features become increasingly critical for longer-term forecasting. Indeed, $\stemmed_{\text{pl}}$ achieves a lower MARE of 0.518 at 1-month ahead and 0.594 at 6-months. Its performance indicates the value of utilizing network and nodal and individual-level information; however, the exponential kernel in $\stemmed$ appears to be more suitable at capturing rapid, short-term triggering effects characteristic of overdose events, which results in its overall superior performance. Overall, these results confirm that while excitation structure is advantageous, the incorporation of time-varying dynamics can further improve OOD forecasting accuracy.
}


{
As for the data-sharing policies, we find that all data-sharing policies perform worse than STEMMED as expected due to limited information sharing. Nonetheless, our experiments reveal that $\stemmed$ under policy $\pi_1$ ($\stemmed_{\pi_1}$) results in the best performance, achieving an MARE of 0.467 at 1-month ahead forecast, which increases to 0.532 at 6-months. This strong performance suggests that drug type-based data-sharing enables $\stemmed$ to maintain high accuracy even with limited information. In contrast, $\stemmed_{\pi_{3,3}}$, which also consists of three information clusters but based on geographic proximity, performs slightly worse. This performance disparity indicates that while spatial proximity captures some dependencies, it may not fully align with the underlying dynamics driving OOD events as effectively as drug type-based information sharing. More restrictive information-sharing policies consistently demonstrate reduced forecast capabilities. For instance, $\stemmed_{\pi_2}$, with its 10 county clusters, shows a higher MARE of 0.521 at 1-month, peaking at 0.59 at 5-months before dropping to 0.554 at 6-months. $\stemmed_{\pi_{3,5}}$, utilizing $5$ distance-based clusters, exhibits even more pronounced performance limitations. These results suggest that increasingly granular geographic clustering can substantially restrict data sharing and impede the model's ability to learn complex spatial relationships.}

}

{
\paragraph{Real-Time OOD Trend Monitoring. 
%
} \label{sec:real_time_monitoring}
In this analysis, we highlight the effectiveness of the STEMMED-based surveillance system with full information for informing the timing of public health interventions by comparing it with a practical evidence-based surveillance approach. We assume that there is a $6$-month lag in data availability for each city-OOD data stream and the cities share a monthly updated database. Moreover, practitioners monitor each city-OOD data stream on a monthly basis using an online cumulative sum (CUSUM) control chart procedure \citep{page1954continuous}. 
We build our STEMMED-based surveillance system using this database and make projections to make up for the data delay. These predictions are subsequently utilized in the CUSUM procedure for detection. 
{
A key objective of our STEMMED-based surveillance system is to detect specific types of changes in OOD patterns to inform the timing of public health interventions. For example, in this analysis, we aim to detect gradual changes in heroin-related OODs, which may indicate a slow resurgence or decline in heroin use, and abrupt changes in fentanyl- and psychostimulant-related OODs, which often signal sudden outbreaks due to the introduction of highly potent fentanyl batches into the drug supply. This is achieved by setting a wider (tighter) decision threshold for heroin (fentanyl and psychostimulant) cases.}
Further details regarding the CUSUM procedure adopted in this study are provided in E-Companion\ref{sec:cusum_procedure} of the E-Companion. 
For comparison, we mirror the procedure with on-time reports, i.e., monthly OOD event data without delays, along with delayed reports. While the former indicates the ideal scenario when OOD data are available for analysis immediately, the latter reflects a realistic scenario in an evidence-based outbreak study.
}

{
\SingleSpacedXI
\begin{figure}[thb]
    \centering
    \begin{minipage}{0.99\textwidth}
    \centering
    \includegraphics[width=0.9\textwidth,keepaspectratio]{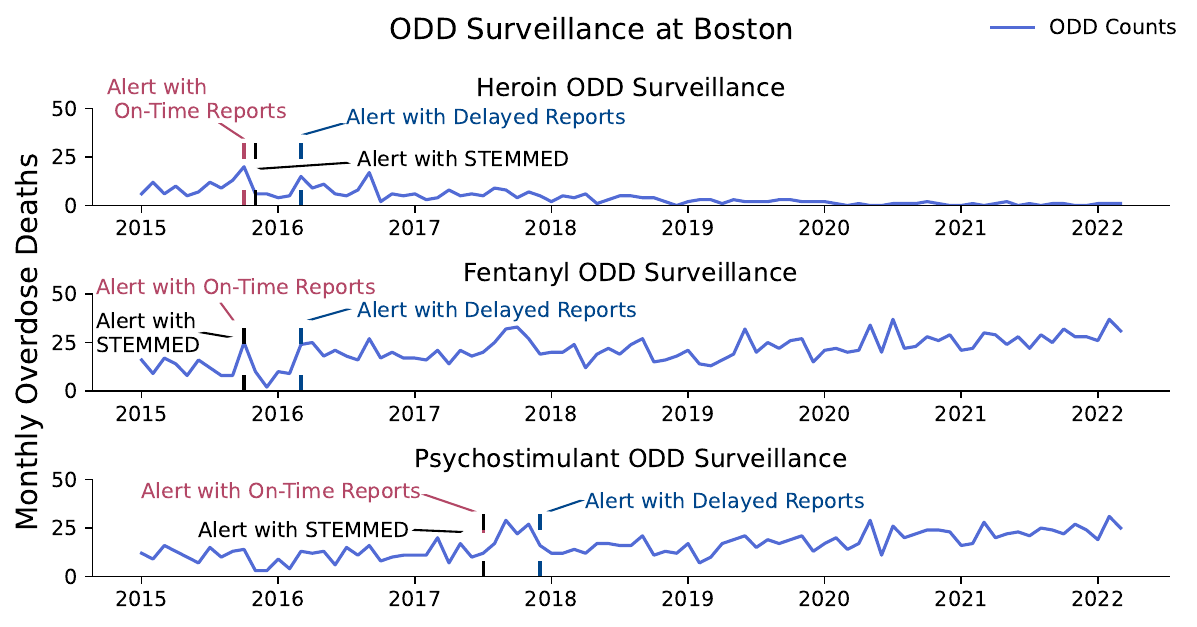}
    \caption{
Change Analysis for Heroin, Fentanyl, and Psychostimulant-Related Overdose Deaths at Boston.
    }
    \label{fig:surveillance_boston}
\end{minipage}
\end{figure}
}

{
 We first provide an illustrative example (Figure \ref{fig:surveillance_boston}) of using the online CUSUM procedure to monitor changes in opioid overdose deaths in Boston, for heroin, fentanyl, and psychostimulant-related cases from January 1, 2015, to April 30, 2022. The impact of delayed reporting was evident, as alerts for changes in overdose deaths due to heroin and fentanyl were delayed until March 2016, and for psychostimulant-related deaths until { end of 2017}. It is noteworthy, however, that the implementation of STEMMED-based surveillance systems reduced these delays significantly. The system improved detection time by 5 months for heroin, { 6 months for fentanyl and for psychostimulant-related cases.} The { average} detection results are summarized in Table \ref{table:detection_delay}, where we define \textit{detection delay} as the time difference between the alerts suggested by on-time reports and those suggested by the STEMMED-based surveillance,
{
i.e., $t_{\stemmed} - t_{\text{ontime}}$, and \textit{\% improvement} as 
$\frac{t_{\text{delayed}} - t_{\stemmed} }{t_{\text{delayed}}}\times100\%$.
}
The results showed an average detection delay of $2.64$ months across all substances, with specific delays of $1.2$ months for heroin, $3.56$ months for fentanyl, and $3.16$ months for psychostimulant overdoses. The $95\%$ confidence intervals were approximately $3$ for most cases, indicating the reliability of these findings. These results underscore the advantages of using the STEMMED-based surveillance system for monitoring OODs. Furthermore, on average, our proposed system achieved a $56\%$ improvement in detection times, compared to delayed reports. In particular, we observe a $80\%$ improvement for heroin-related deaths, $40.67\%$ for fentanyl-related deaths, and $47.33\%$ for psychostimulant-related deaths.}
{
Notably, in heroin cases, the STEMMED-based system can detect changes earlier than on-time reports. Although it could be a false alarm, this result is acceptable because of the small time differences and because the empirical nature of surveillance necessitates further investigation even with the most accurate data reports.
These improvements illustrate the $\stemmed$-based system’s adaptability from forecasting to monitoring of distinct temporal characteristics of different OOD types, which enables more proactive and targeted interventions to address the opioid crisis.}

{
\SingleSpacedXI
\begin{table}[thb]

\centering
\caption{Improvement in Outbreak Monitoring}
\scalebox{0.7}{
\begin{tabular}{|c|l|l|} 
\hline
& Avg. Detection Delay ($95\%$ CI) & Avg. $\%$ Improvement ($95\%$ CI)\\ \hline
All Nodes & $2.64$ $(0.37,\; 4.9)$ & $56\%$ $(18.31\%,\; 93.7\%)$ \\ \hline
Heroin & $1.2\;(-2.04,\; 4.44)
$ & $80\% \;(25.96\%,\; 134\%)
$ \\ \hline
Fentanyl & $3.56\; (2.83,\; 4.28)
$ & $40.67\%\;(28.59\%,\; 52.74\%)
$ \\ \hline
Psychostimulant & $3.16\; (1.12,\; 5.19)
$ & $47.33\% (13.42\%,\; 81.25\%)
$ \\ \hline

\end{tabular}
}
\label{table:detection_delay}
\end{table}
}

{
\subsection{Public Health Policy Implication and Potential Impact}\label{sec:policy_implication}
STEMMED presents a new approach for modeling relationships between cities and illicit OOD events by incorporating both nodal and event-level triggering effects. It creates a network of OOD trends where each trend is defined by its inherent nodal attributes and the triggering effects of OOD events across the network. The model provides policy insights through its transparent framework and precise forecasting capabilities. Notably, STEMMED offers insights into public policy through its clear modeling framework and accurate forecasts. 

Drawing on a multi-year individual-level illicit OOD record and census data in cities in Massachusetts, USA, our analyses supplement the known shift in the opioid crisis with characterizations regarding the decline in heroin-related OODs and the rise in fentanyl- and psychostimulant-driven OOD events. Our results highlight the diminishing inter-drug connections between heroin and other substances and the growing connections between fentanyl and psychostimulants from the parameters, networks, and the triggering effects analyses. Moreover, we also found significant demographic and geographic variations that can potentially inform tailored intervention strategies. Specifically, poverty rate emerges as the most influential nodal feature across all OOD types, which highlights the role of socioeconomic disadvantage in driving OOD risk. At the individual level, age is the dominant triggering factor for heroin- and fentanyl-related OODs, while sex has the largest impact for psychostimulants. Geographically, counties like Norfolk and Suffolk show age- and sex-driven triggering effects. These findings underscore the need for interventions that are sensitive to demographic distinctions. For instance, public health campaigns, outreach, or messaging designed for specific age demographics in Norfolk or sex categories in Suffolk could more effectively address OODs across varying regions and substance types. Additionally, the rapid temporal decay of heroin-related OODs indicates that fentanyl’s longer-lasting effects necessitate sustained surveillance and follow-up intervention, particularly in urban centers like Suffolk where these triggering effects are most pronounced.

Additionally, our online analysis, which simulates the real-time implementation of a STEMMED-based surveillance system, generates promise for the practical use of STEMMED for public health agencies. Specifically, we found that a STEMMED-base surveillance system is remarkably accurate over the established forecasting approaches. Further, this novel surveillance system can reduce the time needed to detect shifts in illicit OOD trends by an $56\%\, (2.6$ months$)$ improvement, compared to the delayed reports which are commonly used in practice \citep{ahmad2022nchs}.

{
In addition, the analysis of data-sharing policies provides policy insights for operationalizing STEMMED in real-world settings. Our analyses indicate that sharing data specific-data can largely preserve $\stemmed$’s predictive accuracy, while geographic-based clustering, i.e., county-based or distance-based can lead to higher errors. These result indicate that overly restrictive collaboration among local entities may hinder the system’s ability to capture spatial relationships. Therefore, prioritizing data-sharing agreements that facilitate drug-type-based collaboration can be promising to enhance surveillance and intervention efforts.}

In sum, our empirical results suggest that STEMMED has the potential to provide a new understanding of OOD trends and aid the subsequent designs of public health policy interventions, including locating precise intervention targets and encouraging multi-city strategic collaboration. Additionally, our research highlight the benefits of adopting a STEMMED-based surveillance system as a means to address the issue of data-processing delays in existing public health systems. {However, we note that similar to other predictive models, actionable insights generated by $\stemmed$ and the model itself can create a feedback loop as actions based on the model may alter outcomes and subsequently reshape the OOD landscape and influence the model performance \citep{lum2016predict}. Thus, it is crucial to continuously monitor performance and adjust the system as needed.}
}
\section{Conclusion} \label{sec: conclusion}
In this work, we introduced a Spatio-TEMporal Mutually Exciting point process with Dynamic network structure ($\stemmed$) to address the needs of an accurate opioid overdose death (OOD) forecasting system that is feasible to implement locally. $\stemmed$ quantifies the subtle connections between past and future OOD events across a wide range of local communities and drug use behaviors and can be easily decomposed node-by-node, which suggests a tractable learning algorithm. Then, using $\stemmed$'s decomposability, we analyze a practical online cooperative forecasting framework which operates at various community levels and 
{study the advantages of multi-location collaboration and data-sharing policies.} 

{
In our case study, we focused on twenty-five cities in Massachusetts with the highest occurrences of illicit OODs and identified a notable cluster around the Boston area. Our findings showed an increasing association over time between fentanyl and psychostimulants. This pattern underscores the necessity for a fentanyl-focused approach in addressing the current OOD trend. Moreover, the $\stemmed$-based surveillance system surpassed other models in predicting both short- and long-term OODs, which demonstrates its proficiency in discerning the complex relationships between historical and future events. Additionally, this system significantly reduced detection delays by more than $60\%$, which enhances the timeliness of interventions based on accurate forecasts to compensate for the delays displayed in current evidence-based surveillance. 
{We also found that drug-based data-sharing may be more advantageous over its geographic-based counterparts.}
}

Our study has limitations. Firstly, causal relationships between the event streams may not be directly inferred from the dynamic network connections in $\stemmed$. Moreover, there might be difficulties for {model estimation} or data sharing between local communities due to {resource} or privacy concerns. Additionally, we assign multi-drug deaths to single drug categories based on city-specific prevalence. This approach preserves marginal distributions of drug types in OODs and ensures computational efficiency, but can lose multi-substance information from an individual OOD. While including all drug combinations could cause computational challenges due to network size and unstable estimation due to data sparsity, future work could explore efficient node pruning strategies to better capture poly-drug interactions. Potentially, a Bayesian model estimation approach could be developed to allow local communities to adopt and update existing models with minimal resources. Future work could also explore computationally efficient alternatives for STEMMED's forecasting, such as using ODE-based methods to compute expected intensities under simplified assumptions, or developing approximation techniques that maintain the model's rich capabilities while reducing simulation costs. Finally, the case study was a single state analysis and $\stemmed$, as a spatio-temporal model, suffers from the so-called boundary effects; that is, the model cannot capture the effects from units outside the modeled network. Large-scale analyses with many participating states may mitigate this issue.

In closing, $\stemmed$ is a novel forecasting tool that captures the dynamic yet subtle links between past and future events of distinct types. While $\stemmed$ was motivated by OOD prediction, its key analytical properties facilitate its application to other domains, including disease surveillance, demand planning, crime analysis, and cybersecurity.

}

\def\bibfont{\scriptsize}
\bibliographystyle{env/informs2014}
\bibliography{reference.bib}




\end{document}